\documentclass[onecolumn,amsmath,amsssymb]{revtex4}
\input epsf
\def\gsim{ \lower .75ex \hbox{$\sim$} \llap{\raise .27ex \hbox{$>$}} }
\def\lsim{ \lower .75ex \hbox{$\sim$} \llap{\raise .27ex \hbox{$<$}} }

\begin{document}

\title{Chameleon Cosmology}

\author{Justin Khoury and Amanda Weltman}

\affiliation{Institute for Strings, Cosmology and Astroparticle Physics, Columbia University, New York, NY 10027}

\begin{abstract}
The evidence for the accelerated expansion of the universe and the time-dependence of the fine-structure constant suggests the existence of at least one scalar field with a mass of order $H_0$. If such a field exists, then it is generally assumed that its coupling to matter must be tuned to unnaturally small values in order to satisfy the tests of the Equivalence Principle (EP). In this paper, we present an alternative explanation which allows scalar fields to evolve cosmologically while having couplings to matter of order unity. In our scenario, the mass of the fields depends on the local matter density: the interaction range is typically of order 1 mm on Earth (where the density is high) and of order $10-10^4$ AU in the solar system (where the density is low).
All current bounds from tests of General Relativity are satisfied.
Nevertheless, we predict that near-future experiments that will test gravity in space will measure an effective Newton's constant different by order unity from that on Earth, as well as EP violations stronger than currently allowed by laboratory experiments. 
Such outcomes would constitute a smoking gun for our scenario.
\end{abstract}

\maketitle

\section{Introduction} \label{intro}

There is growing evidence in cosmology for the existence of nearly massless scalar fields in our Universe. On the one hand, a host of observations, from supernovae luminosity-distance measurements~\cite{sn1a} to the cosmic microwave background anisotropy~\cite{wmap}, suggests that 70\% of the current energy budget consists of a dark energy fluid with negative pressure. While observations are consistent with a non-zero cosmological constant, the dark energy component is more generally modeled as quintessence: a scalar field rolling down a flat potential~\cite{bp,quint}. In order for the quintessence field to be evolving on cosmological time scales today, its mass must be of order $H_0$, the present Hubble parameter.

On the other hand, recent measurements of absorption lines in quasar spectra suggest that the fine-structure constant $\alpha$ has evolved by roughly one part in $10^5$ over the redshift interval $0.2<z<3.7$~\cite{webb}. Time-variation of coupling constants are generally modeled with rolling scalar fields~\cite{bekenstein}, and the recent evidence for a time-varying $\alpha$ requires the mass of the corresponding scalar field to be of order $H_0$~\cite{dvalizal}.

In either case, the inferred scalar field is essentially massless on solar system scales, and therefore subject to tight constraints from tests of the Equivalence Principle (EP)~\cite{willbook}. The current bound on the E$\ddot{{\rm o}}$tv$\ddot{{\rm o}}$s parameter, $\eta$, which quantifies the deviation from the universality of free-fall, is $\eta < 10^{-13}$, from the E$\ddot{{\rm o}}$t-Wash experiment~\cite{eotwash}. 

From a theoretical standpoint, massless scalar fields or moduli are abundant in string and supergravity theories. Indeed, generic compactifications of string theory result in a plethora of massless scalars in the low-energy, four-dimensional effective theory. However, these massless fields generally couple directly to matter with gravitational strength, and therefore lead to unacceptably large violations of the EP. Therefore, if the culprit for quintessence or time-varying $\alpha$ is one of the moduli of string theory, some mechanism must effectively suppress its EP-violating contributions.

For instance, Damour and Polyakov~\cite{pol} (see also~\cite{damnord}) have proposed a dynamical mechanism to suppress the coupling constants $\beta_i$ between the various matter fields and the dilaton of string theory. Alternatively, the suppression could be the result of approximate global symmetries~\cite{sean}.

In a recent paper~\cite{letter}, we presented a scenario in which scalar fields can evolve cosmologically while having couplings to matter of order unity, {\it i.e.,} $\beta_i\sim {\cal O}(1)$. This is because the scalar fields acquire {\it a mass whose magnitude depends on the local matter density}. In a region of high density, such as on Earth, the mass of the fields is large, and thus the resulting violations of the EP are exponentially suppressed. In the solar system, where the density is much lower, the moduli are essentially free, with a Compton wavelength that can be much larger than the size of the solar system. Finally, on cosmological scales, where the density is very low, the mass can be of the order of the present Hubble parameter, thereby making the fields potential candidates for causing the acceleration of the universe or the time-evolution of the fine-structure constant. While the idea of density-dependent mass terms is not new~\cite{pol,carroll,huey,hill}, the novelty of our work lies in that the scalar fields can couple directly to baryons with gravitational strength.

In our scenario, scalar fields that have cosmological effects, such as quintessence, do not result in large violations of the EP in the laboratory because we happen to live in a very dense environment. Thus, the main constraint on our model is that the mass of the field be sufficiently large on Earth to evade EP and fifth force constraints~\cite{fischbach}. 

The generation of a density-dependent mass for a given modulus $\phi$ results from the interplay
of two source terms in its equation of motion. The first term arises from self-interactions, described by a monotonically-decreasing potential $V(\phi)$ which is of the runaway form (see Fig.~\ref{pot}). In particular, we underscore the fact that the potential need not have a minimum; rather, it must be monotonic. The second term arises from the conformal coupling to matter fields, of the form $e^{\beta_i\phi/M_{Pl}}$. The coupling constants $\beta_i$ need not be small, however, and values of order unity or greater are allowed. Although these two contributions are both monotonic functions of $\phi$, their combined effect is that of an effective potential which does display a minimum (see Fig.~\ref{poteff}). Furthermore, since this effective potential depends explicitly on the local matter density $\rho$, both the field value at the minimum and the mass of small fluctuations depend on $\rho$ as well, with the latter being an increasing function of the density.

Although the scalar fields are quite massive on Earth, their behavior is strikingly different in the solar system where the local matter density is much smaller. Thus our model makes a crucial prediction for near-future experiments that will test gravity in space. For example, consider the SEE Project~\cite{SEE} which among other things will measure Newton's constant to an unprecedented accuracy. Our scenario generically predicts that the SEE experiment should observe corrections of order unity to Newton's constant compared to its measured value on Earth, due to fifth-force contributions which are important in space but exponentially suppressed on Earth.

Moreover, three satellite experiments to be launched in the near future, STEP~\cite{STEP}, Galileo Galilei (GG)~\cite{GG} and MICROSCOPE~\cite{MICRO}, will test the universality of free-fall in orbit with expected accuracy of $10^{-18}$, $10^{-17}$ and $10^{-15}$, respectively. We predict that these experiments should observe a strong EP-violating signal. In fact, for a wide range of parameters, our model predicts that the signal will be larger than the ground-based E$\ddot{{\rm o}}$t-Wash bound of $10^{-13}$.

If SEE does measure an effective Newton's constant different from that on Earth, or if STEP observes an EP-violating signal larger than thought permitted by the E$\ddot{{\rm o}}$t-Wash experiment, this will strongly indicate that a mechanism of the form proposed here is realized in Nature. For otherwise it would be hard to explain the discrepancies between measurements in the laboratory and those in orbit. These new and surprising outcomes are a direct consequence of the fact that scalar fields in our model have drastically different behavior in regions of high density than in regions of low density.

We refer to $\phi$ as a ``chameleon'' field, since its physical properties, such as its mass, depend sensitively on the environment. Moreover, in regions of high density, the chameleon ``blends'' with its environment and becomes essentially invisible to searches for EP violation and fifth force.

Even though we predict significant violations of the EP in space, all existing constraints from planetary orbits~\cite{willbook}, such as those from lunar laser ranging~\cite{nord}, are easily satisfied in our model. This is because of the fact that the chameleon-mediated force between two large objects, such as the Earth and the Sun, is much weaker than one would naively expect. To see this, we use calculus and break up the Earth into a collection of infinitesimal volume elements. Consider one such volume element located well-within the Earth. Since the mass of the chameleon is very large inside the Earth, the $\phi$-flux from this volume element is exponentially suppressed and therefore contributes negligibly to the $\phi$-field outside the Earth. This is true for all volume elements within the Earth, except for those located in a thin shell near the surface. Infinitesimal elements within this shell are so close to the surface that they do not suffer from the bulk exponential suppression. Thus, the exterior field is generated almost entirely by this thin shell, whereas the bulk of the Earth contributes negligibly. A similar argument applies to the Sun. Consequently, the chameleon-mediated force between the Earth and the Sun is suppressed by this thin-shell effect, which thereby ensures that solar system tests of gravity are satisfied. 

However, note that this only applies to large objects, such as planets. Sufficiently small objects do not suffer from thin-shell suppression, and thus their entire mass contributes to the exterior field. In particular, a small satellite in orbit, such as SEE, may not exhibit a thin-shell effect. This is why the orbits of the planets are essentially unaffected by the $\phi$-force, whereas the fifth force between two test particles in the SEE capsule is significant.

Since $\phi$ couples directly to matter fields, all mass scales and coupling constants of the Standard Model depend on space and time. Once again due to the thin-shell mechanism described above, spatial variations of constants are sufficiently small in our model to satisfy current experimental bounds, for instance from the Vessot-Levine experiment~\cite{vessot}. Time variation of coupling constants are also not a problem since, during most of the history of the universe, the various couplings actually change by very little. Thus the bounds from big bang nucleosynthesis, for instance, are easily satisfied. This will be described in more detail in a separate paper dealing with the cosmological evolution in our model~\cite{cosmo}.

In Sec.~\ref{setups}, we describe the ingredients of the scenario, focusing on a single modulus $\phi$ for simplicity. We show how the dynamics of $\phi$ are governed by an effective potential that depends on the local matter density. In Sec.~\ref{compact}, we derive approximate solutions for $\phi$ for a compact object such as the Sun, for instance, and describe the thin-shell mechanism mentioned earlier. In Sec.~\ref{earth}, we specialize the solution for $\phi$ to the  case of the Earth, and apply the results in Sec.~\ref{tests} to derive constraints on the parameters of the theory based on laboratory tests of the EP and searches for a fifth force. We then show in Sec.~\ref{condmass} that, for a potential of power-law form, $V(\phi)=M^{4+n}\phi^{-n}$, these constraints translate into the requirement that the energy scale $M$ be less than an inverse millimeter or so. Curiously, this is also the scale associated with the cosmological constant today. In Sec.~\ref{solar}, we argue that our model easily satisfies constraints from solar system tests of GR. It is showed (Sec.~\ref{SEP}) that the same holds true for bounds from spatial and time variation of coupling constants. We then predict (Sec.~\ref{satellite}) that near-future experiments that aim at testing the EP and measuring a fifth force should observe a large signal, perhaps stronger than previously thought possible. Finally, we conclude and summarize our results in Sec.~\ref{conclu}.

\section{The Ingredients of the Model} \label{setups}

Focusing on a single scalar field $\phi$ for simplicity, the action governing the dynamics of our model is given by
\begin{equation}
S=\int d^4x\sqrt{-g}\left\{\frac{M_{Pl}^2}{2}{\cal
R}-\frac{1}{2}(\partial\phi)^2- V(\phi)\right\} 
- \int d^4x{\cal L}_m(\psi_m^{(i)},g_{\mu\nu}^{(i)})\,,
\label{action}
\end{equation}
where  $M_{Pl}\equiv (8\pi G)^{-1/2}$ is the reduced Planck mass, $g$ is the determinant of the metric $g_{\mu\nu}$, ${\cal R}$ is the Ricci
scalar and $\psi_m^{(i)}$ are matter fields. The scalar field $\phi$ interacts directly with matter particles through
a conformal coupling of the form $e^{\beta_i\phi/M_{Pl}}$.
Explicitly, each matter field $\psi_m^{(i)}$ couples
to a metric $g_{\mu\nu}^{(i)}$ which is related to the Einstein-frame metric
$g_{\mu\nu}$ by the rescaling
\begin{equation}
g_{\mu\nu}^{(i)}=e^{2\beta_i\phi/M_{Pl}}g_{\mu\nu}\,,
\label{conformal}
\end{equation}
where $\beta_i$ are dimensionless constants~\cite{dgg}. Moreover, for simplicity, we assume that the different $\psi_m^{(i)}$'s do not interact with each other. Note that Eq.~(\ref{action}) is of the general form of low-energy effective actions from string theory and supergravity, where $V(\phi)$ arises from non-perturbative effects.

\begin{figure}
\epsfxsize=3 in \centerline{\epsfbox{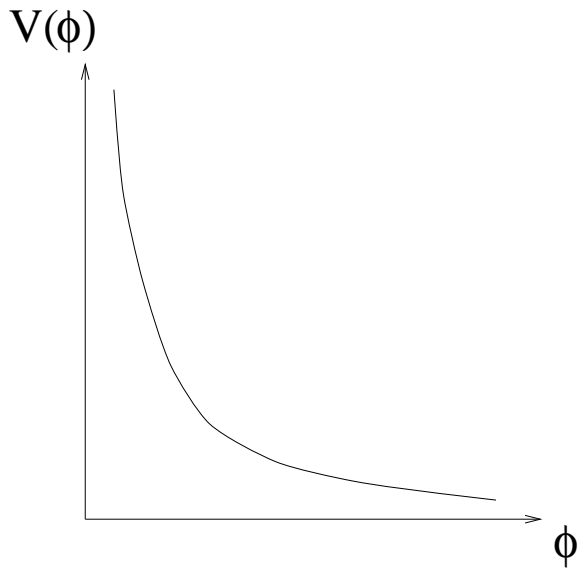}}
\caption{Example of a runaway potential.}
\label{pot}
\end{figure}

The potential $V(\phi)$ is assumed to be of the runaway form. That is, it is monotonically decreasing and satisfies 
\begin{equation}
\lim_{\phi\rightarrow\infty} V =0\,,\qquad \lim_{\phi\rightarrow\infty} \frac{V_{,\phi}}{V} =0\,,
\qquad \lim_{\phi\rightarrow\infty} \frac{V_{,\phi\phi}}{V_{,\phi}} = 0\ldots\,,
\label{Vasym1}
\end{equation}
as well as 
\begin{equation}
\lim_{\phi\rightarrow 0} V =\infty\,,\qquad \lim_{\phi\rightarrow 0} \frac{V_{,\phi}}{V} =\infty\,,
\qquad \lim_{\phi\rightarrow 0} \frac{V_{,\phi\phi}}{V_{,\phi}} = \infty\ldots\,,
\label{Vasym2}
\end{equation}
where $V_{,\phi}\equiv dV/d\phi$, {\it etc}. See Fig.~\ref{pot}.
For instance, a fiducial example is the inverse power-law potential
\begin{equation}
V(\phi)=M^{4+n}\phi^{-n}\,,
\end{equation}
where $M$ has units of mass and $n$ is a positive constant.
The above conditions on the asymptotics of $V$ are generally satisfied by
potentials arising from non-perturbative effects in string
theory~\cite{carroll,huey,hill,rest}.
Note that it is also of the desired form for quintessence models of the universe~\cite{zlatev}.

The equation of motion for $\phi$ derived from the above action is then
\begin{equation}
\nabla^2\phi = V_{,\phi} - \sum_i\frac{\beta_i}{M_{Pl}}e^{4\beta_i\phi/M_{Pl}}g_{(i)}^{\mu\nu}T^{(i)}_{\mu\nu}\,,
\label{eom0}
\end{equation}
where $T^{(i)}_{\mu\nu} = (2/\sqrt{-g^{(i)}})\delta {\cal L}_m/\delta g_{(i)}^{\mu\nu}$ is the stress-energy tensor for the $i$th form of matter. For the purpose of this paper, it will suffice to approximate the geometry as Minkowski space, that is, $g_{\mu\nu}\approx \eta_{\mu\nu}$. This is valid provided that the Newtonian potential is small everywhere, and that the backreaction due to the energy density in $\phi$ is also small. This latter assumption will be justified when we analyze post-Newtonian corrections in Sec.~\ref{PPN}.

For non-relativistic matter, one has $g_{(i)}^{\mu\nu}T^{(i)}_{\mu\nu}\approx -\tilde{\rho}_i$, where $\tilde{\rho}_i$ is the energy density. However, we shall find it convenient to express our equations not in terms of $\tilde{\rho}_i$, but rather in terms of an energy density $\rho_i\equiv \tilde{\rho}_ie^{3\beta_i\phi/M_{Pl}}$ which is conserved in Einstein frame. In other words, $\rho_i$ is defined so that it is independent of $\phi$. Equation~(\ref{eom0}) thus reduces to
\begin{equation}
\nabla^2\phi = V_{,\phi} + \sum_i\frac{\beta_i}{M_{Pl}}\rho_i e^{\beta_i\phi/M_{Pl}}\,.
\label{eom}
\end{equation}

\begin{figure}
\epsfxsize=3 in \centerline{\epsfbox{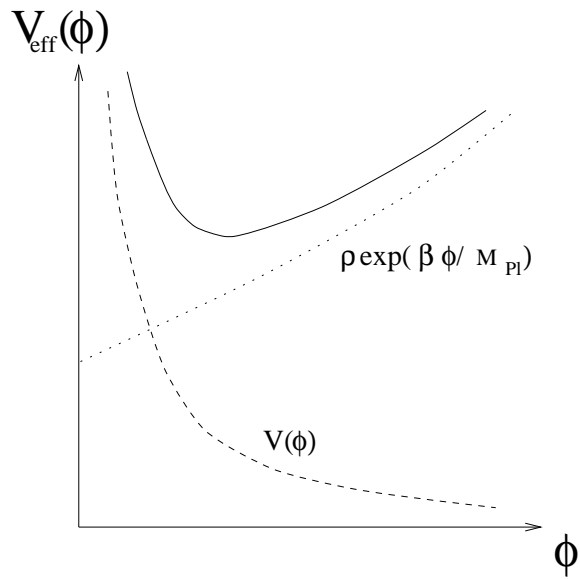}}
\caption{The chameleon effective potential $V_{eff}$ (solid curve) is the sum of two contributions: one from the actual potential $V(\phi)$ (dashed curve), and the other from its coupling to the matter density $\rho$ (dotted curve).}
\label{poteff}
\end{figure}

From the right-hand side of Eq.~(\ref{eom}), we see that the dynamics of
$\phi$ are not solely governed by $V(\phi)$, but rather by an effective potential
\begin{equation}
V_{eff}(\phi) \equiv V(\phi) + \sum_i\rho_i e^{\beta_i\phi/M_{Pl}}\,
\label{veff}
\end{equation}
which depends explicitly on the matter density $\rho_i$. In particular, although
$V(\phi)$ is monotonic, $V_{eff}$ does exhibit a minimum provided that $\beta_i>0$. This is illustrated in Fig.~\ref{poteff} for the case of a single component $\rho$ with coupling $\beta$. (One could equivalently consider the case $V_{,\phi}>0$ and $\beta_i<0$.) Unfortunately, known examples in string theory have $V_{,\phi}$ and $\beta_i$ occurring with the same sign. If $\phi$ is the modulus of an extra dimension, for instance, then one expects that $V\rightarrow 0$ and $V_{,\phi}\rightarrow 0$ in the decompactification limit~\cite{dineseiberg}; in this limit, the various masses of particles also tend to zero.

We will denote by $\phi_{min}$ the value assumed by $\phi$ at the minimum, that is,
\begin{equation}
V_{,\phi}(\phi_{min}) + \sum_i\frac{\beta_i}{M_{Pl}}\rho_i e^{\beta_i\phi_{min}/M_{Pl}}=0\,.
\label{phimin}
\end{equation}
Meanwhile, the mass of small fluctuations about $\phi_{min}$ is obtained
as usual by evaluating the second derivative of the potential at
$\phi_{min}$:
\begin{equation}
m_{min}^2 = V_{,\phi\phi}(\phi_{min}) + \sum_i\frac{\beta_i^2}{M_{Pl}^2}\rho_i
e^{\beta_i\phi_{min}/M_{Pl}}\,.
\label{mmin}
\end{equation}

\begin{figure}
\epsfxsize=5 in \centerline{\epsfbox{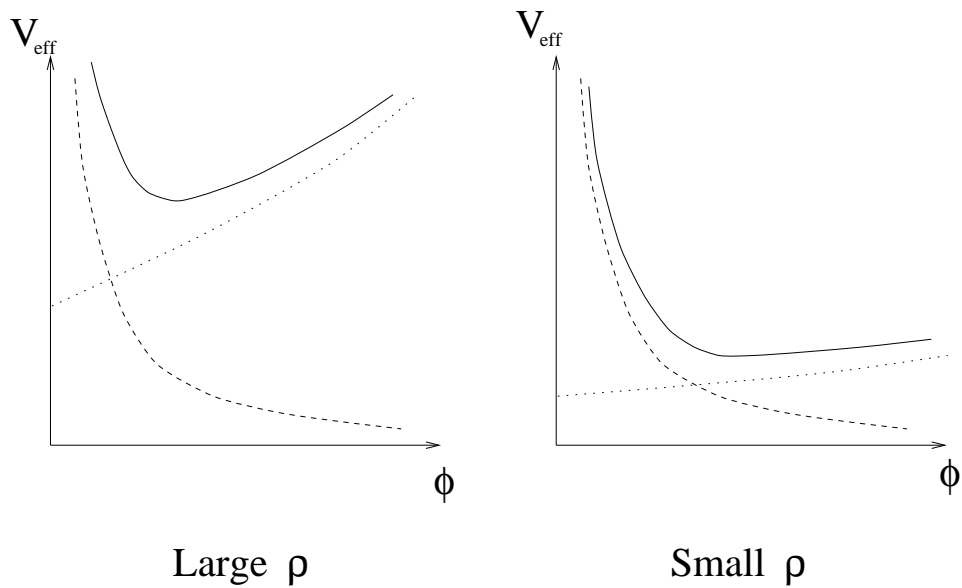}}
\caption{Chameleon effective potential for large and small $\rho$, respectively. This illustrates that, as $\rho$ decreases, the minimum shifts to larger values of $\phi$ and the mass of small fluctuations decreases. (Line styles are the same as in Fig.~\ref{poteff}.)}
\label{lrhosrho}
\end{figure}

Equations~(\ref{phimin}) and~(\ref{mmin}) respectively imply that the local value of the field, $\phi_{min}$, and its mass, $m_{min}$,
both depend on the local matter density.
Since $V_{,\phi}$ is negative and monotonically increasing, while
$V_{,\phi\phi}$ is positive and decreasing,
it also follows that larger values of
$\rho_i$ correspond to smaller $\phi_{min}$ and larger $m_{min}$.
This is illustrated in Fig.~\ref{lrhosrho}. The denser the environment, the more massive is the chameleon.
We will later see that it is possible for $m_{min}$ to be sufficiently large on Earth to evade current constraints on EP violations and fifth force, while being sufficiently small on cosmological scales for $\phi$ to have interesting cosmological effects.

The upshot of our model, from a theoretical standpoint, is that the potential $V(\phi)$ need not have a minimum, nor does the coupling constant $\beta$ need be tuned to less than $10^{-4}$ to satisfy EP constraints~\cite{pol}. Quite the contrary, $V(\phi)$ is assumed monotonic, while $\beta$ can be of order unity. 

\section{Profile for a Compact Object} \label{compact}

In order to study the observable consequences of our model, in particular with regards to EP violations and fifth force mediation, we must first understand the profile that $\phi$ acquires on Earth and in the solar system. Therefore, in this Section, we derive an approximate solution for $\phi$ in the case where the source is a compact object, which we idealize as being perfectly spherical and having homogeneous density. 

Thus consider a static, spherically-symmetric body of radius $R_c$, homogeneous density $\rho_c$ and total mass $M_c=4\pi R_c^3\rho_c/3$.
In the case of the Earth, the latter can be approximated by the characteristic terrestrial density:
$\rho_\oplus \approx 10\; {\rm g}/{\rm cm}^3$. We treat the object as isolated, in the sense that the effect of surrounding bodies is neglected.
It is not in vacuum, however, but is instead immersed in a background of homogeneous density
$\rho_\infty$. In the case of solar system objects, this models the fact that our local neighborhood of the galaxy is not empty, but rather filled with an approximately homogeneous component of baryonic gas and dark matter with density $\rho_\infty\equiv\rho_G\approx 10^{-24}\;{\rm g}/{\rm cm}^3$. In the case of a baseball in the Earth's atmosphere, $\rho_\infty$ denotes the surrounding atmospheric density: $\rho_\infty\equiv\rho_{atm}\approx 10^{-3}\;{\rm g}/{\rm cm}^3$. 

With these assumptions, Eq.~(\ref{eom}) reduces to
\begin{equation}
\frac{d^2\phi}{dr^2} + \frac{2}{r}\frac{d\phi}{dr} = V_{,\phi} +
\frac{\beta}{M_{Pl}}\rho (r) e^{\beta\phi/M_{Pl}}\,,
\label{sun1}
\end{equation}
where
\begin{equation}
\rho (r) = \left\{
\begin{matrix}
\rho_c\qquad {\rm for}\;\;\; r<R_c \cr
\rho_\infty\qquad {\rm for}\;\;\; r>R_c
\end{matrix}
\right. \,.
\label{rho}
\end{equation}
Note that we temporarily focus on the case where all $\beta_i$'s assume the same value $\beta$.
This is done for simplicity only, and the following analysis remains
qualitatively unchanged when these are taken to be
different.
Moreover, this assumption will be dropped when we derive the resulting
violations of the EP in Sec.~\ref{tests}.
In other words, in the end we are not assuming that the theory is Brans-Dicke~\cite{bd}.

Throughout the analysis, we denote by $\phi_c$ and $\phi_\infty$ the field value which minimizes $V_{eff}$ for $r<R_c$ and $r>R_c$, respectively.
That is, from Eqs.~(\ref{phimin}) and~(\ref{rho}), we have
\begin{eqnarray}
\nonumber
& & V_{,\phi}(\phi_c) + \frac{\beta}{M_{Pl}}\rho_c e^{\beta\phi_c/M_{Pl}} = 0\;; \\
& & V_{,\phi}(\phi_\infty) + \frac{\beta}{M_{Pl}}\rho_\infty e^{\beta\phi_\infty/M_{Pl}} = 0\,.
\end{eqnarray}
Similarly, we denote by $m_c$ and $m_\infty$ the mass of small fluctuations about $\phi_c$ and $\phi_\infty$, respectively. That is, $m_c$ ($m_\infty$) is the mass of the chameleon field inside (outside) the object.

Equation~(\ref{sun1}) is a second order differential equation and as such requires two boundary conditions.
Since the solution must be non-singular at the origin, we require $d\phi/dr=0$ at $r=0$, as usual. Moreover, since $\rho = \rho_\infty$ at infinity, it is natural to impose $\phi \rightarrow \phi_\infty$ as $r\rightarrow\infty$. This latter condition is physically sensible as it implies $d\phi/dr\rightarrow 0$ as $r\rightarrow\infty$, and thus that the $\phi$-force between the compact object and a test particle tends to zero as their separation becomes infinite. To summarize, Eq.~(\ref{sun1})
is subject to the following two boundary conditions:
\begin{eqnarray}
\nonumber
& & \frac{d\phi}{dr} = 0 \qquad {\rm at}\;\;\;r=0\,; \\
& & \phi\rightarrow \phi_\infty \qquad {\rm as}\;\;\; r\rightarrow \infty\,.
\label{bc}
\end{eqnarray}

\subsection{Qualitative description of the solution} \label{descrip}

Before solving this problem explicitly, it is useful to give a heuristic derivation of the solution. Far outside the object, $r\gg R_c$, we know that the chameleon tends to $\phi_\infty$, as required by the second boundary condition. There are then two types of solution, depending on whether the object is large ({\it e.g.}, the Earth) or small ({\it e.g.}, a baseball). The distinction between large and small will be made precise below. 

Small objects do not generate large variations in $\phi$. Thus, their solution can be thought of as a perturbation on the background solution $\phi=\phi_\infty$. Hence, one has $\phi\approx\phi_\infty$ everywhere in this case, including the interior of the object.

Large objects, on the other hand, are strongly perturbing. Within the object, $r<R_c$, one finds that the chameleon nearly minimizes the effective potential, and thus $\phi\approx\phi_c$. Hence, the solution essentially extrapolates between $\phi=\phi_c$ within the core and $\phi=\phi_\infty$ far outside.

\begin{figure}
\epsfxsize=3 in \centerline{\epsfbox{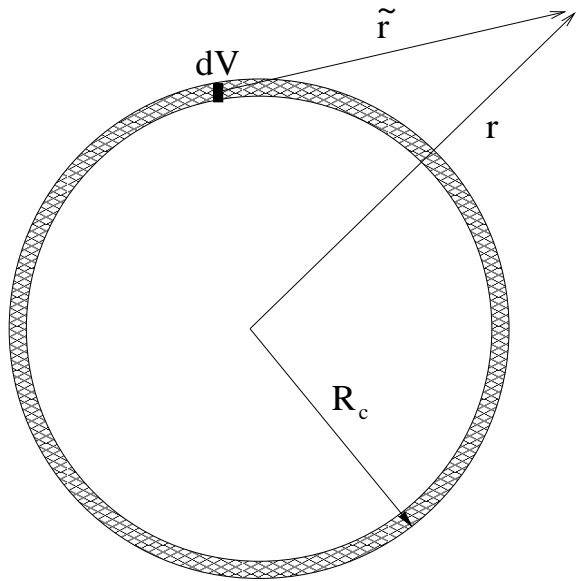}}
\caption{For large objects, the $\phi$-field a distance $r>R_c$ from the center is to a good approximation entirely determined by the contribution from infinitesimal volume elements $dV$ (dark rectangle) lying within a thin shell of thickness $\Delta R_c$ (shaded region). This thin-shell effect suppresses the resulting chameleon force.}
\label{neman}
\end{figure}

To be more precise, let us describe the exterior solution ($r>R_c$) for large bodies, assuming $m_\infty R_c\ll 1$ for simplicity. For this purpose, it is convenient to break up the object into infinitesimal volume elements $dV$ and consider their individual contribution to the $\phi$-field. Well-within the object, one has $\phi\approx\phi_c$, and the mass of the chameleon is large, $m_c\gg m_\infty$. Thus, the contribution from a volume element $dV$ within the core is proportional to $\exp(-m_c\tilde{r})$ and is therefore exponentially suppressed. In other words, it contributes negligibly to the $\phi$-field outside. This holds for all infinitesimal volume elements within the object, except for those lying within a thin shell of thickness $\Delta R_c$ near the surface~\cite{nemancomment}. See Fig.~\ref{neman}. Thus, the exterior solution is obtained by summing over all elements within this shell:
\begin{equation}
\phi(r)\approx -\left(\frac{\beta}{4\pi M_{Pl}}\right)\left(\frac{3\Delta R_c}{R_c}\right)\frac{M_ce^{-m_\infty r}}{r} + \phi_\infty\,,
\label{thinsoln}
\end{equation}
where $r$ is the distance from the center of the object. 
We will find in Sec.~\ref{rigor} that
\begin{equation}
\frac{\Delta R_c}{R_c} = \frac{\phi_\infty-\phi_c}{6\beta M_{Pl}\Phi_c}\,,
\label{DR}
\end{equation}
where $\Phi_c=M_c/8\pi M_{Pl}^2R_c$ is the Newtonian potential at the surface of the object. The derivation of Eq.~(\ref{thinsoln}) implicitly assumed that the shell was thin, that is,
\begin{equation}
\frac{\Delta R_c}{R_c}\ll 1\,.
\label{thincond}
\end{equation}
We shall henceforth refer to this as the thin-shell condition.

Small objects, in the sense that $\Delta R_c/R_c>1$, do not have a thin shell. Rather, their entire volume contributes to the $\phi$-field outside, and thus the exterior solution is
\begin{equation}
\phi(r)\approx -\left(\frac{\beta}{4\pi M_{Pl}}\right)\frac{M_ce^{-m_\infty r}}{r} + \phi_\infty\,,
\label{warmup}
\end{equation}
which is recognized as the Yukawa profile for a scalar field of mass $m_\infty$. Note that Eq.~(\ref{thinsoln}) and~(\ref{warmup}) differ only by a thin-shell suppression factor of $\Delta R_c/R_c$. To summarize, the exterior solution for a compact object is given by 
\begin{eqnarray}
\nonumber
& & \phi(r)\approx -\left(\frac{\beta}{4\pi M_{Pl}}\right)
\frac{M_ce^{-m_\infty r}}{r} + \phi_\infty 
\qquad \;\;\;\;\;\;\;\;\;\;\;\;\;\;\;\;{\rm if}\;\;\; \frac{\Delta R_c}{R_c} > 1\;; \\
& & \phi(r)\approx -\left(\frac{\beta}{4\pi M_{Pl}}\right)
\left(\frac{3\Delta R_c}{R_c}\right)
\frac{M_ce^{-m_\infty r}}{r} + \phi_\infty \qquad {\rm if}\;\;\; \frac{\Delta R_c}{R_c} \ll 1\,,
\label{phisummary}
\end{eqnarray}
with $\Delta R_c/R_c$ defined in Eq.~(\ref{DR}).

The ratio $(\phi_\infty-\phi_c)/M_{Pl}\Phi_c$, which appears in Eq.~(\ref{DR}) and determines whether or not an object has a thin shell, can be interpreted physically as follows. Given a background profile $\phi(r)$, it is straightforward to show from the action~(\ref{action}) that the resulting chameleon-force on a test particle of mass $M$ and coupling $\beta$ is given by
\begin{equation}
\vec{F}_\phi=-\frac{\beta}{M_{Pl}} M\vec{\nabla}\phi\,.
\label{fifth}
\end{equation}
It follows that $\phi$ should be thought of as a potential for this ``fifth'' force. Thus, $(\phi_\infty-\phi_c)/M_{Pl}\Phi_c$ is the ratio of the difference in $\phi$ potential to the Newtonian potential, and effectively quantifies how perturbing the object is for the $\phi$ field.

\subsection{Derivation} \label{rigor}

\begin{figure}
\epsfxsize=5 in \centerline{\epsfbox{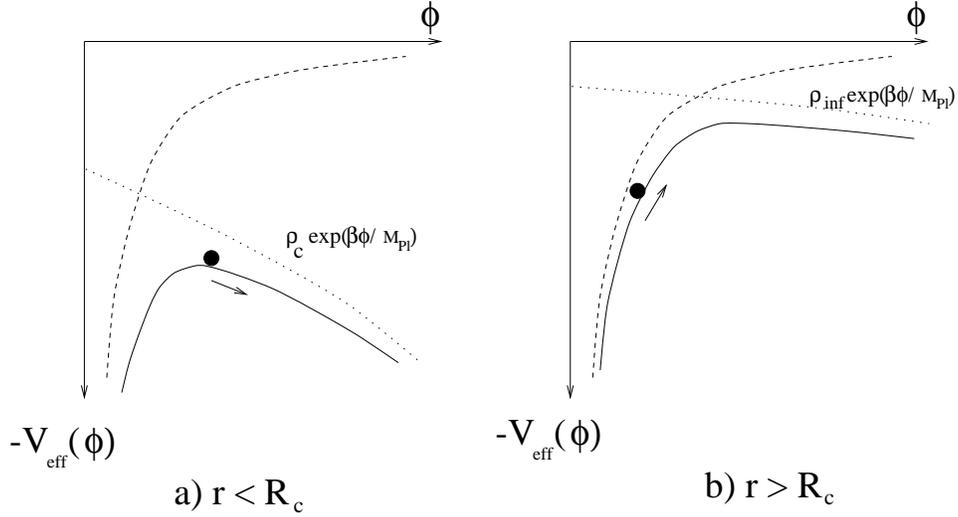}}
\caption{The inverted potential $-V_{eff}$ for a compact object of radius $R_c$ is discontinuous at $r=R_c$ since the matter density equals: a) $\rho_c$ for $r<R_c$; b) $\rho_\infty$ for $r>R_c$. The dots represent the position of the particle at some value of $r$.}
\label{potr}
\end{figure}

To get an intuition for the boundary value problem at hand, it is useful to think of $r$ as a time coordinate and $\phi$ as the position of a particle, and treat Eq.~(\ref{sun1}) as a dynamical problem in classical mechanics. This is akin to the familiar trick performed in bubble nucleation calculations~\cite{coleman}. In this language, the particle moves along the inverted potential $-V_{eff}$, and the second term on the left-hand side of Eq.~(\ref{sun1}), proportional to $1/r$, is recognized as a damping term. An important difference here is that $-V_{eff}$ is ``time''-dependent since $\rho$ depends on $r$. More precisely, the effective potential undergoes a jump at time $r=R_c$, as illustrated in Fig.~\ref{potr}.

The particle begins at rest, since $d\phi/dr=0$ at $r=0$, from some initial value which we denote by $\phi_i$:
\begin{equation}
\phi_i \equiv \phi(r=0)\,.
\end{equation}
For small $r$, the friction term is large, and thus the particle is essentially frozen at $\phi=\phi_i$. It remains stuck there until the damping term, proportional to $1/r$, is sufficiently small to allow the driving term, $dV_{eff}/d\phi$, to be effective. In other words, the amount of ``time'' the particle remains stuck near $\phi=\phi_i$ depends on the slope of the potential, $dV_{eff}/d\phi$, at $\phi=\phi_i$. Once friction is negligible, the particle begins to roll down the potential. See Fig.~\ref{potr}a).

It rolls down until, at some later time $r=R_c$, the potential suddenly changes shape as $\rho(r)$ undergoes a jump from $\rho_c$ to $\rho_\infty$ (see Eq.~(\ref{rho})). But $\phi$ and $d\phi/dr$ are of course continuous at the jump, and the particle keeps rolling, this time climbing up the inverted potential. See Fig.~\ref{potr}b). If the initial position $\phi_i$ is carefully chosen, the particle will barely reach $\phi_\infty$ in the limit $r\rightarrow\infty$, as desired. (It is easy to prove that such $\phi_i$ always exists.) Thus the problem is reduced to determining the initial value $\phi_i$. 

In the end $\phi_i$ will depend on the physical properties of the compact object, such as its density $\rho_c$ and its radius $R_c$, as well as on the various parameters of the theory, such as $\beta$ and the shape of the potential. Rather than first choosing a set of values for $\rho_c$, $R_c$, {\it etc.} and then solving for $\phi_i$, we will instead choose a range of $\phi_i$ and determine the corresponding region in the $(\rho_c,R_c,\ldots)$ parameter space. More precisely, we shall consider the two regimes $(\phi_i-\phi_c)\ll \phi_c$ and $\phi_i\;\gsim\;\phi_c$, and we will show that these correspond respectively to $\Delta R_c/R_c\ll 1$ and $\Delta R_c/R_c > 1$. Anticipating this result, we refer to these two regimes as {\it thin-shell} and {\it thick-shell}, respectively.

\noindent {\bf Thin-shell regime: $(\phi_i-\phi_c)\ll\phi_c$.} This corresponds to $\phi$ being released from a point very close to $\phi_c$. Since $\phi_c$ is a local extremum of the effective potential, the driving term $dV_{eff}/d\phi$ is negligible initially, and the dynamics are strongly dominated by friction. Consequently, the field remains frozen at its initial value $\phi_i\approx \phi_c$ for a long time, until the friction force is sufficiently small to allow the particle to roll. We shall denote by $R_{roll}$ the ``moment'' at which this occurs. Hence, we have
\begin{equation}
\phi(r)\approx \phi_c\qquad{\rm for}\;\;\; 0<r<R_{roll} \,.
\label{frozen}
\end{equation}
When $r\sim R_{roll}$, the field is still near $\phi_c$ but has now begun to roll. Since $M_{Pl}|V_{,\phi}|\ll \beta\rho e^{\beta\phi/M_{Pl}}$ as soon as $\phi$ is displaced significantly from $\phi_c$, as illustrated in Fig.~\ref{poteff}, we may approximate Eq.~(\ref{sun1}) in the regime $R_{roll}<r<R_c$ by
\begin{equation}
\frac{d^2\phi}{dr^2} + \frac{2}{r}\frac{d\phi}{dr} \approx
\frac{\beta}{M_{Pl}}\rho_c\,,
\label{approx1}
\end{equation}
where we have also assumed $\beta\phi/M_{Pl}\ll 1$. The solution to Eq.~(\ref{approx1}) with boundary conditions $\phi=\phi_c$ and $d\phi/dr=0$ at $r=R_{roll}$ is
\begin{equation}
\phi(r) = \frac{\beta\rho_c}{3M_{Pl}}\left(\frac{r^2}{2}+\frac{R_{roll}^3}{r}\right)-
\frac{\beta\rho_c R_{roll}^2}{2M_{Pl}} + \phi_c\qquad{\rm for}\;\;\;R_{roll}<r<R_c\,.
\label{rlrc2}
\end{equation}
The full solution for $0<r<R_c$ is thus approximated by Eqs.~(\ref{frozen}) and~(\ref{rlrc2}). The approximation of separating the solution into the two regions $0<r<R_{roll}$ and $R_{roll}<r<R_c$ only makes sense, however, if $R_c-R_{roll}\ll R_c$. For otherwise there is no clear separation between the two regions, and one needs a solution valid over the entire range $0<r<R_c$.

At $r=R_c$, the energy density undergoes a jump from $\rho_c$ to $\rho_\infty$, as
described by Eq.~(\ref{rho}). For $r>R_c$, the effective potential is shown in Fig.~\ref{potr} b),
and the particle is climbing up the hill. Since the speed of the particle is initially large
compared to the curvature of the potential, Eq.~(\ref{sun1}) can be approximated by
\begin{equation}
\frac{d^2\phi}{dr^2} + \frac{2}{r}\frac{d\phi}{dr} \approx 0\,,
\end{equation}
whose solution, satisfying $\phi\rightarrow \phi_\infty$ as $r\rightarrow\infty$, is given by
\begin{equation}
\phi(r) \approx -\frac{Ce^{-m_\infty (r-R_c)}}{r} + \phi_\infty \,,
\label{better2}
\end{equation}
where $C$ is a constant and where we have used the fact that the potential is approximately quadratic near $\phi=\phi_\infty$.

The two unknowns, $R_{roll}$ and $C$, are then determined by matching $\phi$ and $d\phi/dr$ at $r=R_c$ using Eqs.~(\ref{rlrc2}) and~(\ref{better2}). With the approximation that $R_c-R_{roll}\ll R_c$, it is straightforward to show that the exterior solution is 
\begin{equation}
\phi(r)\approx -\left(\frac{\beta}{4\pi M_{Pl}}\right)
\left(\frac{3\Delta R_c}{R_c}\right)
\frac{M_ce^{-m_\infty (r-R_c)}}{r} + \phi_\infty\,,
\label{rgrc2fin}
\end{equation}
with
\begin{equation}
\frac{\Delta R_c}{R_c}\equiv \frac{\phi_\infty-\phi_c}{6\beta M_{Pl}\Phi_c} \approx \frac{R_c-R_{roll}}{R_c} \ll 1\,,
\label{Rrollsmall}
\end{equation}
where we have substituted the Newtonian potential $\Phi_c$. 

\noindent {\bf Thick-shell regime: $\phi_i\;\gsim\;\phi_c$.} In this case, the field is initially sufficiently displaced from $\phi_c$ that it begins to roll almost as soon
as it is released at $r=0$. Hence there is no friction-dominated regime in this case,
and the interior solution for $\phi$ is most easily obtained by taking the $R_{roll}\rightarrow 0$ limit of Eq.~(\ref{rlrc2}) and replacing $\phi_c$ by $\phi_i$. Matching to the exterior solution as before, we obtain
\begin{equation}
\phi(r)=\frac{\beta\rho_cr^2}{6 M_{Pl}} + \phi_i\qquad{\rm for}\;\;\;0<r<R_c
\end{equation}
and
\begin{equation}
\phi(r)\approx -\left(\frac{\beta}{4\pi M_{Pl}}\right)\frac{M_ce^{-m_\infty (r-R_c)}}{r} + \phi_\infty \qquad{\rm for}\;\;\;r>R_c\,.
\label{sol3}
\end{equation}
Moreover, equating these two equations at $r=R_c$, we find $\phi_i = \phi_\infty - 3\beta\Phi_c/M_{Pl}$. In particular, since $\phi_i\;\gsim\;\phi_c$ and using the definition $\Delta R_c/R_c = (\phi_\infty-\phi_c)/6\beta M_{Pl}\Phi_c$, this implies
\begin{equation}
\frac{\Delta R_c}{R_c}> 1\,.
\end{equation}

We conclude this Section with a word on how the above solutions, which assumed homogeneous $\rho$,
can be generalized to the more realistic case of spatially-varying matter density.
In most cases of interest, such as the interior of the Earth for instance, we will find that
the matter density varies on scales much larger than the Compton wavelength $m^{-1}$ of the chameleon field
in that region. More precisely, it is generally the case that $|\nabla\log\rho(\vec{x})|\ll m$ within dense objects.
If so, one can make an adiabatic approximation which consists of treating $\rho(\vec{x})$ as a constant
in the equations of motion. In other words, in this case one may simply substitute $\rho(\vec{x})$ in the
expressions above.

\section{Profile for the Earth} \label{earth}

Since the most stringent constraints on possible violations of the EP derive from experiments performed on Earth, it is important to discuss in some detail the profile for $\phi$ inside and in the vicinity of the Earth. Admittedly, the model for our planet described below is rather crude, but is sufficiently accurate to derive order-of-magnitude estimates of resulting
violations of the EP. More realistic descriptions can be obtained for instance by using the adiabatic approximation discussed at the end of Sec.~\ref{compact}, or through numerical analysis.

The Earth is modeled as a solid sphere of radius $R_{\oplus}=6\cdot 10^8$ cm and homogeneous density $\rho_{\oplus}=10\;{\rm g}/{\rm cm}^3$. Surrounding it is an atmosphere which we approximate as a layer 10 km in radius with homogeneous density $\rho_{atm} = 10^{-3}\;{\rm g}/{\rm cm}^3$. Moreover, we treat our planet as an isolated body, neglecting the effect of surrounding compact objects such as the Sun and the Moon. Furthermore, far away from the Earth, the matter density is approximated by the density of homogeneous gas and dark matter in our local neighborhood of the galaxy: $\rho_{G} = 10^{-24}\;{\rm g}/{\rm cm}^3$.

The set-up is thus almost identical to that of Eq.~(\ref{sun1}), except that the matter density now has three phases instead of two: 
\begin{equation}
\rho (r) = \left\{
\begin{matrix}
\;\;\;\rho_\oplus\qquad\;\;\; {\rm for}\;\;\; 0<r<R_\oplus \cr
\;\;\;\;\;\;\;\;\;\;\rho_{atm}\qquad {\rm for}\;\;\; R_\oplus< r <R_{atm} \cr
\rho_G\qquad\;\;\; {\rm for}\;\;\; r>R_{atm}
\end{matrix}
\right. \,,
\label{rhoearth}
\end{equation}
where $R_{atm}\equiv R_\oplus +$ 10 km. We henceforth denote by $\phi_\oplus$, $\phi_{atm}$ and $\phi_G$ the field value which minimizes the effective potential for the respective densities. Similarly, $m_\oplus$, $m_{atm}$ and $m_G$ are the respective masses.

Following the discussion in Sec.~\ref{compact}, the solution depends on whether or not the Earth and its atmosphere have a thin shell. As we will prove below, it is necessary that the atmosphere has a thin shell, for otherwise unacceptably large violations of the EP will ensue. In this case, one has $\phi\approx \phi_{atm}$ in the bulk of the atmosphere. Moreover, since the Earth is much denser than the atmosphere, it follows that the Earth itself has a thin shell, in which case $\phi\approx\phi_\oplus$ inside the Earth.

From Eqs.~(\ref{DR}) and~(\ref{thincond}), the thin-shell condition for the atmosphere reads
\begin{equation}
\frac{\Delta R_{atm}}{R_{atm}} = \frac{\phi_G-\phi_{atm}}{6\beta M_{Pl}\Phi_{atm}} \ll 1\,,
\end{equation}
where $\Phi_{atm}\equiv \rho_{atm}R_{atm}^2/6M_{Pl}^2$. We can refine this bound by noting that, in order for the atmosphere to have a thin shell, clearly the thickness of the shell must be less than the thickness of the atmosphere itself, which is $\approx 10^{-3} R_{atm}$. Hence this requires $\Delta R_{atm}/R_{atm} \;\lsim\; 10^{-3}$. Using the fact that $\rho_{atm}\approx 10^{-4}\rho_{\oplus}$, and thus $\Phi_{atm}\approx 10^{-4}\Phi_\oplus$, we can write this as
\begin{equation}
\frac{\Delta R_\oplus}{R_\oplus}\equiv \frac{\phi_G-\phi_{atm}}{6\beta M_{Pl}\Phi_{\oplus}} < 10^{-7} \,.
\label{condatm}
\end{equation}
This condition, which ensures that the atmosphere has a thin shell, will play a crucial role in the analysis of tests of gravity in the following sections. The exterior solution, $r>R_{atm}$, is then given by Eq.~(\ref{rgrc2fin}) with $m_\infty=m_G$ and $\phi_\infty=\phi_G$:
\begin{equation}
\phi(r) \approx -\left(\frac{\beta}{4\pi M_{Pl}}\right)
\left(\frac{3\Delta R_\oplus}{R_\oplus}\right)
\frac{M_\oplus e^{-m_G(r-R_{atm})}}{r} + \phi_G \,.
\label{earthext}
\end{equation}

To summarize, the solution for $\phi$ for the Earth and its atmosphere is well-approximated by
\begin{eqnarray}
\nonumber
& & \phi(r) \approx \phi_{\oplus} \qquad\;\;\;\;\;\;\;\;\;\;\;\;\;\;\;\;\;\;\;\;\;\;\;\;\;\;\;\;\;\;\;\;\;\;\;\;\;\;\;\;\;\;\;\;\;\;\;\;\;\;\;\;\;\;\;\;\;\;\;\;\;\;\;\;\;\;\;\;\;\;{\rm for}\;\;\;0 < r \;\lsim\; R_{\oplus}
\;; \\
\nonumber
& & \phi(r) \approx \phi_{atm} \qquad\;\;\;\;\;\;\;\;\;\;\;\;\;\;\;\;\;\;\;\;\;\;\;\;\;\;\;\;\;\;\;\;\;\;\;\;\;\;\;\;\;\;\;\;\;\;\;\;\;\;\;\;\;\;\;\;\;\;\;\;\;\;\;\;\;\;\;{\rm for}\;\;\; R_{\oplus} \;\lsim\; r \;\lsim\;
R_{atm} \;; \\
& & \phi(r) \approx -\left(\frac{\beta}{4\pi M_{Pl}}\right)
\left(\frac{3\Delta R_\oplus}{R_\oplus}\right)
\frac{M_\oplus e^{-m_G(r-R_{atm})}}{r} + \phi_G \qquad{\rm for}\;\;\; r\;\gsim\; R_{atm} \,,
\label{earthsummary}
\end{eqnarray}
where $\Delta R_\oplus/R_\oplus$ is defined in Eq.~(\ref{condatm}).

It remains to show that tests of the EP require the atmosphere to have a thin shell. The proof proceeds by contradiction. Suppose that condition~(\ref{condatm}) is violated and instead we have
\begin{equation}
\frac{\Delta R_\oplus}{R_\oplus} > 10^{-7}\,.
\label{condatmviolated}
\end{equation}
We can therefore ignore the atmosphere altogether, and thus the $\phi$-profile in the laboratory is given by Eq.~(\ref{earthext}):
\begin{equation}
\phi(r) \approx -\left(\frac{\beta}{4\pi M_{Pl}}\right)
\left(\frac{3\Delta R_\oplus}{R_\oplus}\right)
\frac{M_\oplus}{r} + \phi_G \qquad{\rm for}\;\;\; r\;\gsim\; R_\oplus\,,
\end{equation}
where we have neglected the exponential factor since $m_G R_\oplus\ll 1$, as we will see in Sec.~\ref{condmass}. From Eq.~(\ref{fifth}), this profile results in a fifth force on a test particle of mass $M$ and coupling $\beta_i$ of magnitude
\begin{equation}
|\vec{F}_\phi| = 2\beta\beta_i 
\left(\frac{3\Delta R_\oplus}{R_\oplus}\right)
\frac{M_{\oplus}M}{8\pi M_{Pl}^2r^2}\,.
\end{equation}
Supposing that the $\beta_i$'s are all of order $\beta$ but assume different values for different matter species, then the resulting difference in relative free-fall acceleration for two bodies of different composition will be
\begin{equation}
\eta \equiv 2\frac{|a_1 - a_2|}{a_1+a_2} \sim 10^{-4}\beta^2\frac{\Delta R_\oplus}{R_\oplus}\,,
\end{equation}
where $\eta$ is the E$\ddot{{\rm o}}$tv$\ddot{{\rm o}}$s parameter, and where the numerical coefficient is appropriate for Cu and Be test masses~\cite{pol}, as used in the E$\ddot{{\rm o}}$t-Wash experiment~\cite{eotwash}. For $\beta$ of order unity, we see from Eq.~(\ref{condatmviolated}) that this violates the bound $\eta<10^{-13}$. It follows that the atmosphere must be have a thin shell.

\section{Searches for EP Violation and Fifth Force on Earth}
\label{tests}

The tightest constraints on our model derive from laboratory tests of the EP and searches for a fifth force~\cite{willbook,fischbach}. Since these experiments are usually done in vacuum, we first need to derive an approximate solution for the chameleon inside a vacuum chamber. For simplicity, we model the chamber as a perfectly empty, spherical cavity of radius $R_{vac}$. In the absence of any other parts within the chamber, and ignoring the effect of the walls, the equation for $\phi$ is given by Eq.~(\ref{sun1}):
\begin{equation}
\frac{d^2\phi}{dr^2} + \frac{2}{r}\frac{d\phi}{dr} =  V_{,\phi}
+\frac{\beta}{M_{Pl}}\rho (r) \,,
\label{vacc}
\end{equation}
where we have assumed $\beta\phi/M_{Pl} \ll 1$ as we did throughout Sec.~\ref{compact}, and where
\begin{equation}
\rho (r) \approx \left\{
\begin{matrix}
0 \qquad \;\;\;\;\;{\rm for}\;\;\; r<R_{vac} \cr
\rho_{atm}\qquad {\rm for}\;\;\; r>R_{vac}
\end{matrix}
\right. \,.
\label{rhovac}
\end{equation}
The boundary conditions are the same as in Sec.~\ref{compact}: $d\phi/dr=0$ at $r=0$ and $\phi \rightarrow
\phi_{atm}$ as $r\rightarrow\infty$. 

The solution within the vacuum chamber is analogous to the solution for a compact object with thin-shell (see Sec.~\ref{compact}). In both cases, due to the large density contrast between the object or the vacuum cavity and their environment, $\phi$ must start at $r=0$ from a point where it can remain almost frozen for the entire volume. That is, for $0<r<R_c$ in the case of the overdense object and $0<r<R_{vac}$ for the vacuum cavity. For a compact object, this freezing point lies naturally near
the local extremum $\phi=\phi_c$ of the effective potential. For the vacuum chamber, the effective potential has no extremum for $r<R_{vac}$ (since $\rho=0$ there), and thus the only way $\phi$ can remain still is by starting from a point where it will be friction-dominated for almost
the entire range $0<r<R_{vac}$. In other words the curvature of the potential at that
point must be of order $R_{vac}^{-2}$, that is, this freezing point corresponds to a value $\phi=\phi_{vac}$ where the mass of small fluctuations, $m_{vac}$, is equal to $R_{vac}^{-1}$.

While the precise solution to~(\ref{vacc}) depends of course on the details of the potential, numerical analysis confirms the above qualitative discussion:
\begin{itemize}
\item The chameleon assumes the value $\phi\sim \phi_{vac}$ within the vacuum chamber, where $\phi_{vac}$ satisfies
\begin{equation}
m_{vac} \equiv \sqrt{V_{,\phi\phi}(\phi_{vac})} = R_{vac}^{-1}\,.
\label{mv}
\end{equation}
That is, $\phi_{vac}$ is the field value about which the Compton wavelength
of small fluctuations equals $R_{vac}$, the radius of the chamber.
\item Throughout the chamber, $\phi$ varies slowly, with $|d\phi/dr|\; \lsim\; \phi_{vac}/R_{vac}$.
\item Outside the chamber the solution tends to $\phi_{atm}$ within a distance of $m_{atm}^{-1}$ from the walls.
\end{itemize}
\noindent These generic properties are all we need for our analysis.

\subsection{Fifth Force Searches} 

The potential energy associated with fifth force interactions is generally parameterized by a Yukawa potential:
\begin{equation}
V(r) =  - \alpha \frac{M_1M_2}{8\pi M_{Pl}^2}\frac{e^{-r/\lambda}}{r}\,,
\label{fifthpot}
\end{equation}
where $M_1$ and $M_2$ are the masses of two test bodies, $r$ is their
separation, $\alpha$ is the strength of the interaction
(with $\alpha=1$ for gravitational strength), and $\lambda$ is the range. Null
fifth-force searches therefore constrain regions in the $(\lambda,\alpha)$ parameter space (see Fig.~2.13 of~\cite{fischbach}).

As discussed above, the range $\lambda$ of $\phi$-mediated interactions inside a
vacuum chamber is of the order of the size of the chamber. That is, $\lambda \approx R_{vac}$. 
For $\lambda\approx 10\;{\rm cm}-1\;{\rm m}$, the tightest bound on the coupling
constant $\alpha$ from laboratory experiments is from Hoskins {\it et al.}~\cite{hoskins}: 
\begin{equation}
\alpha < 10^{-3}\,. 
\label{hos}
\end{equation}

Now consider two identical test bodies of uniform density $\rho_c$, radius $R_c$ and total mass $M_c$. If these have no thin shell, then they each generate a field profile given by Eq.~(\ref{warmup}) with $\phi_\infty = \phi_{vac}$ and $m_\infty = R_{vac}^{-1}$:
\begin{equation}
\phi(r)\approx -\left(\frac{\beta}{4\pi}\right)\frac{M_c}{M_{Pl}}\frac{e^{-r/R_{vac}}}{r}  + \phi_{vac} \,.
\end{equation}
Dropping the irrelevant constant, the resulting potential energy is
\begin{equation}
V(r) = -2\beta^2\frac{M_c^2}{8\pi M_{Pl}^2}\frac{e^{-r/R_{vac}}}{r} \,.
\end{equation}
Comparison with Eq.~(\ref{fifthpot}) shows that the coupling strength is $\alpha = 2\beta^2$ in this case, which clearly violates the bound in Eq.~(\ref{hos}) for $\beta\sim {\cal O}(1)$. 

Hence it must be that the test masses have a thin shell, that is, 
\begin{equation}
\frac{\Delta R_c}{R_c}\equiv \frac{\phi_{vac}-\phi_c}{6\beta M_{Pl}\Phi_c} \ll 1 \,.
\label{condbodyvac}
\end{equation}
In this case, their field profile is given by Eq.~(\ref{thinsoln}):
\begin{equation}
\phi(r)\approx -\left(\frac{\beta}{4\pi M_{Pl}}\right)\left(\frac{3\Delta R_c}{R_c}\right)\frac{M_ce^{-r/R_{vac}}}{r} + \phi_{vac}\,,
\label{bodyvac}
\end{equation}
and the corresponding potential energy is
\begin{equation}
V(r) = -2\beta^2\left(\frac{3\Delta R_c}{R_c}\right)^2
\frac{M_c^2}{8\pi M_{Pl}^2}\frac{e^{-r/R_{vac}}}{r} \,.
\end{equation}
Once again comparing with Eq.~(\ref{fifthpot}), we find that the bound in Eq.~(\ref{hos}) translates into
\begin{equation}
2\beta^2\left(\frac{3\Delta R_c}{R_c}\right)^2 \;\lsim\; 10^{-3}\,.
\label{condbodyvac2}
\end{equation}
Note that, $\beta\;\gsim \; {\cal O}(1)$, this constraint implies that the thin-shell condition in Eq.~(\ref{condbodyvac}) is satisfied.

To make the condition~(\ref{condbodyvac2}) more explicit, note that a typical test body used in Hoskins {\it et al.} had mass $M_c\approx 40$~g and radius $R_c\approx 1$~cm, corresponding to $\Phi_c\approx 3\cdot 10^{-27}$. Substituting in Eq.~(\ref{condbodyvac}) and assuming $\phi_{vac} \gg \phi_c$, we obtain the constraint
\begin{equation}
\phi_{vac} \; \lsim \; 10^{-28}\;M_{Pl}\,,
\label{condvacfin}
\end{equation}
which ensures that the current bounds from laboratory searches of a fifth force are satisfied. 

\subsection{Tests of the EP} 

Turning our attention to the magnitude of EP violations inside our vacuum cavity, we recall that the solution for $\phi$ inside the chamber satisfies
\begin{equation}
\left\vert\frac{d\phi}{dr}\right\vert\; \lsim \;\frac{\phi_{vac}}{R_{vac}} \,.
\label{boundgrad}
\end{equation}
From Eq.~(\ref{fifth}), this yields an extra component to the acceleration of magnitude $(\beta/M_{Pl})\phi_{vac}/R_{vac}$. For Cu and Be test masses, as used in the E$\ddot{{\rm o}}$t-Wash experiment~\cite{eotwash}, this yields an E$\ddot{{\rm o}}$tv$\ddot{{\rm o}}$s parameter of
\begin{equation}
\eta \sim 4\pi 10^{-4}\beta\frac{M_{Pl}R_\oplus^2}{M_\oplus}\frac{\phi_{vac}}{R_{vac}}\,.
\label{boundeta}
\end{equation}
Substituting $R_{vac}=10$ cm and $\beta\sim {\cal O}(1)$, the E$\ddot{{\rm o}}$t-Wash constraint of $\eta<10^{-13}$ translates into
\begin{equation}
\phi_{vac} \;\lsim\; 10^{-26}\;M_{Pl}\,,
\label{condvacfin2}
\end{equation}
which is a much weaker constraint than Eq.~(\ref{condvacfin}) and thus shall henceforth be ignored.

\section{Resulting Constraints on Model Parameters} \label{condmass}

In this Section we summarize the constraints derived in the previous sections and apply them to a general power-law potential
\begin{equation}
V(\phi) = M^{4+n}\phi^{-n}\,,
\label{Vtypical}
\end{equation}
where $M$ has units of mass and $n$ is a positive constant. As mentioned earlier, potentials of
this form have the desired features for quintessence models of the universe~\cite{zlatev}. We will find that the energy scale $M$ is generally constrained to be of the order of $(1\; {\rm mm})^{-1}$. We then discuss the resulting bounds on the interaction range in the atmosphere, in the solar system and on cosmological scales.

Broad considerations of EP violation lead us to conclude in Sec.~\ref{earth} that both the Earth and its atmosphere must have a thin shell. This resulted in the constraint (see Eq.~(\ref{condatm})):
\begin{equation}
\frac{\Delta R_\oplus}{R_\oplus} \equiv \frac{\phi_G-\phi_{atm}}{6\beta M_{Pl}\Phi_{\oplus}} < 10^{-7}\,.
\label{condition2}
\end{equation}
Recall that, by definition, $\phi_G$ is the value of $\phi$ which minimizes the effective potential with $\rho=\rho_G$, {\it i.e.}, $V_{,\phi}(\phi_G) + \beta\rho_Ge^{\beta\phi_G/M_{Pl}}/M_{Pl}=0$. Assuming $\beta\phi_G/M_{Pl}\ll 1$ and substituting the power-law potential of Eq.~(\ref{Vtypical}) gives
\begin{equation}
\phi_G = \left(\frac{nM^{4+n}M_{Pl}}{\beta\rho_G}\right)^{1/(n+1)}\,.
\label{phiGsolved}
\end{equation}
With $\rho_G = 10^{-24}\;{\rm g}/{\rm cm}^3$ and $\Phi_\oplus=10^{-9}$, it is then straightforward to show that Eq.~(\ref{condition2}) can be rewritten as a bound on $M$:
\begin{equation}
M\;<\;\left(\frac{6^{n+1}}{n}\right)^{\frac{1}{n+4}}\beta^{\frac{n+2}{n+4}}\cdot 10^{\frac{15n-7}{n+4}}\cdot (1\;{\rm mm})^{-1}\,.
\label{condM1}
\end{equation}

Then, in Sec.~\ref{tests}, we studied laboratory tests of the EP and fifth force, including the fact that these are performed in vacuum, and derived the condition
\begin{equation}
\phi_{vac} \; \lsim \; 10^{-28}\;M_{Pl}\,,
\end{equation}
where $\phi_{vac}$ is the field value about which the Compton wavelength of small fluctuations is of the order of the size of the vacuum chamber, $R_{vac}$. Applying Eq.~(\ref{mv}) to the power-law potential yields
\begin{equation}
R_{vac}^{-2} = 10^{-68}\;M_{Pl}^2= n(n+1) M^{4+n}\phi_{vac}^{-(n+2)} \,,
\label{Rv}
\end{equation}
where we have assumed $R_{vac}=1\;{\rm m} = 10^{34}\;M_{Pl}^{-1}$ for concreteness. It is easy to see
that smaller values of $R_{vac}$ result in weaker constraints, and this is therefore a conservative choice. It is then straightforward to show that Eq.~(\ref{condvacfin}) reduces
\begin{equation}
M\; \lsim\; [n(n+1)]^{-1/(4+n)} \cdot 10^{3n/(4+n)} \cdot (1\;{\rm mm})^{-1}\,.
\label{condM2}
\end{equation}

We thus see that, for $n$ and $\beta$ of order unity, Eqs.~(\ref{condM1}) and~(\ref{condM2}) both constrain $M$ to be
less than approximately an inverse millimeter, or $10^{-3}$ eV. It is remarkable
that this is also the mass scale associated with the cosmological
constant or dark energy causing the present acceleration of the universe~\cite{cosmo}.
Granted, this is tiny compared to typical particle physics scales, such
as the weak or the Planck scale, and thus our potential~(\ref{Vtypical}) suffers from
fine-tuning. Nevertheless, it is our hope that whatever mechanism
suppresses the scale of the cosmological constant from its natural value
of $10^{18}$ GeV down to $10^{-3}$ eV might also naturally account for the energy
scale characterizing our potential.

The above constraints can be translated into bounds on the range of chameleon-mediated interactions
in the atmosphere ($m_{atm}^{-1}$), in the solar system ($m_G^{-1}$) and on cosmological scales today ($m_0^{-1}$). 
From Eq.~(\ref{mmin}), these are given by
\begin{eqnarray}
\nonumber
& & m_{atm}^2 = V_{,\phi\phi}(\phi_{atm}) + \frac{\beta^2}{M_{Pl}^2}\rho_{atm} e^{\beta\phi_{atm}/M_{Pl}} \\
\nonumber
& & m_G^2 = V_{,\phi\phi}(\phi_G) + \frac{\beta^2}{M_{Pl}^2}\rho_G e^{\beta\phi_G/M_{Pl}}\\
& & m_0^2 = V_{,\phi\phi}(\phi_0) + \frac{\beta^2}{M_{Pl}^2}\rho_0 e^{\beta\phi_0/M_{Pl}} \,,
\end{eqnarray}
where $\rho_0\approx  10^{-29}\;{\rm g}/{\rm cm}^3$ is the current energy density of the universe and $\phi_0$ is the corresponding value of $\phi$ on cosmological scales.
Substituting the above bounds on $M$, it is straightforward to show that, for $n\;\lsim \;2$ and $\beta$ of order unity, 
\begin{eqnarray}
\nonumber
& & m_{atm}^{-1} \;\lsim\; 1\;{\rm mm}-1\;{\rm cm} \\
\nonumber
& & m_G^{-1} \;\lsim\; 10-10^4\;{\rm AU} \\
& & m_0^{-1} \;\lsim\;  0.1-10^3\;{\rm pc}\,,
\label{masses}
\end{eqnarray}
where the numbers on the right-hand depend somewhat on the value of $n$ and $\beta$ (recall that $1\;{\rm AU}\approx 1.5\cdot 10^{13}$ cm and $1\;{\rm pc} \approx 3\cdot 10^{18}$ cm).

Thus, while $\phi$-interactions are short-range in the atmosphere (short compared to the size of the atmosphere), they are rather long-range in the solar system. In particular, it is possible for the scalar field to be essentially free on solar system scales. This striking difference in behavior of the field in space compared to Earth is an original ingredient of our scenario and, as we will show in Sec.~\ref{satellite}, can lead to unexpectedly large signals for EP and fifth force experiments to be performed in orbit in the near future. On cosmological scales, we see that a power-law potential implies an interaction range which is smaller than the present size of the observable universe, $H_0^{-1}\sim 10^{9}\;{\rm pc}$. It follows that $m_0$ is too large for $\phi$ to be rolling on cosmological time scales today. A general study of potentials and their viability for quintessence models of the universe will be presented in a future paper~\cite{cosmo}.

\section{Solar System Tests of Gravity} \label{solar}

In this Section, we discuss the constraints on scalar-tensor theories from planetary orbits, in particular from solar system tests of the EP and fifth force (Sec.~\ref{solarEP}), as well as post-Newtonian corrections (Sec.~\ref{PPN}). In short, these constraints are all satisfied because large bodies, such as the Sun and the Earth, all have a thin shell which greatly suppresses the $\phi$-force between them.

\subsection{Solar System Tests of EP and Fifth Force} \label{solarEP}

Precise measurements of the lunar orbit from laser ranging constrains the difference in  free-fall acceleration of the Moon and 
the Earth towards the Sun to be less than approximately one part in $10^{13}$~\cite{willbook}.
That is, denoting their respective acceleration by $a_{Moon}$ and $a_\oplus$, we have
\begin{equation}
\frac{|a_{Moon} - a_\oplus|}{a_N} \;\lsim\; 10^{-13}\,,
\label{diffacc}
\end{equation}
where $a_N$ is the Newtonian acceleration.

In our model, this difference is naturally very small since the Sun, Earth and Moon are all subject to the thin shell effect. We have already imposed that the Earth (and its atmosphere) have a thin shell. So must the Sun, therefore, since its Newtonian potential is larger than that of the Earth. Hence we only need to show that the same holds true for the Moon. But, assuming $\phi_G\gg \phi_{Moon}$, this trivially follows from Eq.~(\ref{condatm}):
\begin{equation}
\frac{\Delta R_{Moon}}{R_{Moon}}\sim \frac{\Delta R_\oplus}{R_\oplus}\frac{\Phi_\oplus}{\Phi_{Moon}} < 10^{-5}\,,
\end{equation}
where we have used $\Phi_\oplus=10^{-9}$ and $\Phi_{Moon}=10^{-11}$.

Hence the $\phi$ profile outside each of these bodies is given by Eq.~(\ref{rgrc2fin}) with $m_\infty=m_G$ and $\phi_\infty=\phi_G$. Assuming $m_G^{-1}\;>\;1\;{\rm AU}$, since this yields maximal violation of the EP, it is then straightforward to show that the acceleration of the Earth towards the Sun is given by
\begin{equation}
a_\oplus = a_N\cdot\left\{1+18\beta^2\left(\frac{\Delta R_\oplus}{R_\oplus}\right)\left(\frac{\Delta R_\odot}{R_\odot}\right)\right\}\approx a_N\cdot\left\{1+18\beta^2\left(\frac{\Delta R_\oplus}{R_\oplus}\right)^2\frac{\Phi_\oplus}{\Phi_\odot}\right\} \,,
\end{equation}
while for the Moon
\begin{equation}
a_{Moon} = a_N\cdot\left\{1+18\beta^2\left(\frac{\Delta R_{Moon}}{R_{Moon}}\right)\left(\frac{\Delta R_\odot}{R_\odot}\right)\right\} \approx a_N\cdot\left\{1+18\beta^2\left(\frac{\Delta R_\oplus}{R_\oplus}\right)^2\frac{\Phi_\oplus^2}{\Phi_\odot\Phi_{Moon}}\right\}\,.
\end{equation}
Substituting $\Phi_\odot=10^{-6}$, $\Phi_\oplus=10^{-9}$ and $\Phi_{Moon}=10^{-11}$, this gives a difference in free-fall acceleration of 
\begin{equation}
\frac{|a_{Moon} - a_\oplus|}{a_N} \approx \beta^2 \left(\frac{\Delta R_\oplus}{R_\oplus}\right)^2 < \beta^2\cdot 10^{-14}\,,
\end{equation}
where we have used Eq.~(\ref{condatm}) in the last step. This satisfies the bound in Eq.~(\ref{diffacc}) for reasonable values of $\beta$.

We next consider solar-system tests of the existence of a fifth force.
Deviations from a $1/r^2$ force law, for instance due to the exponential factor in Eq.~(\ref{fifthpot}),
contribute an anomalous component to the perihelion precession of planetary orbits in comparison
with the predictions of GR. For instance, Lunar laser-ranging measurements lead to the 
constraint $\alpha\;\lsim\;10^{-10}$ for a fifth-force with range $\lambda\sim 10^8\; {\rm m}$~\cite{dickey}.
A similar analysis for the orbits of Mercury and Mars gives $\alpha\;\lsim\;10^{-9}$ for the range 
$\lambda\sim 1$ AU~\cite{talmadge}. (Not surprisingly, these tests are most sensitive to a fifth force whose range is of the order of the distance between the Sun and the orbiting body.) 
In our model, these celestial objects are all subject to the thin-shell effect, and, just as with the EP analysis above, the
screening mechanism makes the constraints from perihelion precession trivial to satisfy.

\subsection{Tests of Post-Newtonian Gravity} \label{PPN}

To estimate the constraints from post-Newtonian corrections, consider the $\phi$-profile due to the Earth given by Eq.~(\ref{earthext}):
\begin{equation}
\phi(r) \approx -\left(\frac{\beta}{4\pi M_{Pl}}\right)\left(\frac{3\Delta R_\oplus}{R_\oplus}\right)\frac{M_\oplus}{r} + \phi_G\,,
\end{equation}
where we have neglected the exponential factor. Comparison with the expected profile if there were no thin-shell suppression, given by Eq.~(\ref{sol3}), we see that the exterior solution above corresponds to that of a massless scalar with effective coupling
\begin{equation}
\beta_{eff} = 3\beta \frac{\Delta R_\oplus}{R_\oplus}< 3\beta\cdot 10^{-7}\,,
\label{betaeff}
\end{equation}
where in the last step we have used the condition that the atmosphere has a thin-shell (see Eq.~(\ref{condatm})). Treating our model as a Brans-Dicke theory with effective coupling constant $\beta_{eff}$ given above, which is a good approximation in the solar system since the chameleon behaves essentially as a free field, it is straightforward to show that the corresponding effective Brans-Dicke parameter, $\omega_{BD}$, is given by~\cite{willbook}
\begin{equation}
3+2\omega_{BD} = \frac{1}{2\beta_{eff}^{2}} \;\gsim\; 6\cdot 10^{12}\beta^{-2}\,.
\end{equation}
The tightest constraint on Brans-Dicke theories comes from light-deflection measurements using very-long-baseline radio interferometer~\cite{willbook}: $\omega_{BD} > 3500$. We see that this is easily satisfied in our model. Similarly, one can show that the constraint from the decay of the orbital period of binary pulsars, $\omega_{BD}> 100$, is trivially satisfied. (Note that density-dependent effective couplings were previously noted in a different context~\cite{gef}.)

\section{Tests of the Strong EP} \label{SEP}

Our discussion of the EP has so far been restricted to the weak EP, which essentially states that the laws of gravity are the same in any inertial frame. General Relativity, however, satisfies a stronger version of the EP~\cite{willbook}, in the sense that all laws of physics, including non-gravitational interactions, assume the same form in any inertial frame (local Lorentz invariance), and the various parameters describing these non-gravitational forces, such as the fine-structure constant, are independent of space and time (local position invariance). 

At any space-time event, one can find a coordinate system in which the Einstein-frame metric $g_{\mu\nu}$ equals the Minkowski metric $\eta_{\mu\nu}$. Since all other metrics $g_{\mu\nu}^{(i)}$ appearing in the action~(\ref{action}) are conformally-related to $g_{\mu\nu}$, all $g_{\mu\nu}^{(i)}$ are proportional to $\eta_{\mu\nu}$ in this frame. All laws of physics are therefore Lorentz invariant at that space-time event, and thus our model satisfies local Lorentz invariance.

The most stringent constraint on spatial variations of couplings comes from the Vessot-Levine experiment~\cite{vessot} which measured the redshift
between a hydrogen-maser clock flown at an altitude of $10^4$~km and another one on the ground.
In a theory which has LPI, as in GR, the redshift $z$ is given by the difference in Newtonian potential $\Delta\Phi$
between the emitter and the receiver~\cite{wald}. As shown in Will~\cite{willbook}, LPI violations generate an
extra contribution $\Delta z$ to the redshift, of the form
\begin{equation}
\Delta z= \gamma\cdot\Delta\Phi\,,
\end{equation}
where $\gamma$ is a constant that depends on the Newtonian potential of the emitter, with $\gamma=0$ corresponding
to the case of no LPI violation. The bound from the Vessot-Levine experiment is $|\gamma|\;\lsim\; 10^{-4}$.

To estimate $\gamma$ in our case, recall from Eq.~(\ref{conformal}) that a test particle of the matter field
$\psi_m^{(i)}$ follows geodesics of the metric $g_{\mu\nu}^{(i)}$ related to the Einstein-frame metric $g_{\mu\nu}$
by $g_{\mu\nu}^{(i)}=e^{2\beta_i\phi/M_{Pl}}g_{\mu\nu}$. Thus, a constant mass scale $m^{(i)}$ in the $\psi_m^{(i)}$-frame is related to a
$\phi$-dependent mass scale $m(\phi)$ in Einstein frame by the rescaling $m(\phi) = e^{\beta_i\phi/M_{Pl}}m^{(i)}$. 
Similarly, a $\psi_m^{(i)}$-clock with frequency $\nu^{(i)}$ is measured in Einstein frame to have a frequency $\nu(\phi) = e^{\beta_i\phi/M_{Pl}}\nu^{(i)}$. 

Assuming $\beta\phi/M_{Pl}\ll 1$, the $\phi$-dependence of $\nu$ therefore yields the following extra contribution to the redshift:
\begin{equation}
\Delta z \approx -\frac{\beta}{M_{Pl}}(\phi(r_{em})-\phi(r_{rec}))\,,
\label{delz1}
\end{equation}
where we have dropped the superscript $(i)$, and where $r_{em}$ ($r_{rec}$) is the distance between the emitter
(receiver) and the center of the Earth. For the Vessot-Levine experiment, we have
$r_{em}\approx 10^4\;{\rm km}\;\approx 2R_\oplus$ and $r_{rec}\;\gsim\; R_\oplus$, and it follows from
Eqs.~(\ref{earthsummary}) that $\phi(r_{em})\approx \phi_G$ and $\phi(r_{rec})\approx\phi_{atm}$, respectively. Since $\Delta\Phi\approx \Phi_\oplus/2$ in this case, massaging Eq.~(\ref{delz1}) gives
\begin{equation}
\Delta z = -12\beta^2\left(\frac{\phi_G-\phi_{atm}}{6\beta M_{Pl}\Phi_\oplus}\right)\Delta\Phi\,,
\end{equation}
from which we can read off the corresponding value of $\gamma$:
\begin{equation}
|\gamma| = 12\beta^2\frac{\Delta R_\oplus}{R_\oplus} \;\lsim\; \beta^2\cdot 10^{-6} \,.
\end{equation}
This comfortably satisfies the Vessot-Levine bound of $|\gamma|\;\lsim\;10^{-4}$.

Time-variation of coupling constants are constrained by geophysical measurements (such as the Oklo Nuclear Reactor~\cite{oklo}), by the study of absorption lines in quasar spectra and by nucleosynthesis. Recent analysis indicates that the fine-structure constant has evolved by more than one part in $10^5$ over the cosmological redshift range $0.2 < z< 3.7$~\cite{webb}, thus suggesting that LPI does not hold in the Universe.

In our scenario, the various coupling constants are determined by $\phi$, whose value depends on the local density.
Thus, even though the coupling constants may vary on cosmological scales today, the fact that the density of the Earth is constant in time implies that the various couplings measured on Earth do not vary significantly. In particular, this implies that constraints from the Oklo Nuclear Reactor are easily evaded in our model. (See~\cite{mota} for a related discussion of a density-dependent fine-structure constant.) 

In any case, we find that the time-variation of coupling constants and masses on cosmological scales is very small in the case of the power-law potential of Eq.~(\ref{Vtypical}). 
To see this, consider once again a constant mass scale $m^{(i)}$ in the matter frame, with corresponding $m(\phi) = e^{\beta_i\phi/M_{Pl}}m^{(i)}$ in the Einstein frame. Hence, the time variation of $m(\phi)$ between nucleosynthesis and the present epoch, say, is given by
\begin{equation}
\left\vert\frac{\Delta m}{m}\right\vert \approx \frac{\beta}{M_{Pl}} (\phi_0 - \phi_{BBN})\,,
\end{equation}
where $\phi_{BBN}$ is the value of $\phi$ at nucleosynthesis. For the power-law potential considered earlier, $V(\phi) = M^{4+n}\phi^{-n}$, the bound of $M\;\lsim\; (1\;{\rm mm})^{-1}$ derived in Sec.~\ref{condmass} gives
\begin{equation}
\left\vert\frac{\Delta m}{m}\right\vert\;\lsim \; \beta\cdot 10^{-11}\,,
\end{equation}
which satisfies all current astrophysical and cosmological bounds. In particular, it appears that the power-law potential cannot account for the variation of the fine-structure constant reported by Webb {\it et al.}~\cite{webb}. A more detailed analysis of time-variation of coupling constants in our model, including more general potentials, will be presented elsewhere~\cite{cosmo}.

\section{New Predictions for Near-Future Satellite Experiments} \label{satellite}

Our scenario has the remarkable feature that the physical characteristics of the scalar field can be very different in the laboratory than in space. We have seen, for instance, that the range of the interactions it mediates is of order 1~mm in the atmosphere while being greater than 10~AU in the solar system. Hence, as we
will show in this Section, the predicted strength of EP violations and fifth force can be strikingly
different in orbit than in the laboratory. This is particularly interesting as this decade should witness the 
launch of several satellite experiments that aim at making these measurements. Three of them, namely STEP, GG and
MICROSCOPE, will test the universality of free-fall with respective expected sensitivity of $10^{-18}$, $10^{-17}$ and
$10^{-15}$ for the E$\ddot{{\rm o}}$tv$\ddot{{\rm o}}$s parameter $\eta$. Another satellite, the SEE Project, will
further attempt to measure a fifth force between two test bodies in orbit. 

In this Section we will show that, for a wide range of parameters, our model predicts that these satellite
experiments will see a strong signal. It is in fact likely that the signal will be stronger than previously
thought possible based on laboratory measurements. For instance, the SEE capsule could detect corrections to the
effective Newton's constant (including fifth force contribution) of order unity compared to the value measured on Earth or
inferred from planetary motion. Moreover, STEP, GG and MICROSCOPE could measure violations of the EP with
$\eta > 10^{-13}$, which is larger than the current bound from the (ground-based) E$\ddot{{\rm o}}$t-Wash experiment. Such outcomes would constitute a smoking gun for our model, for it would otherwise be difficult to reconcile the results in space
with those on Earth.

Let us begin with the SEE satellite, which will orbit the Earth at an altitude of approximately $10^3$ km.
This multi-faceted experiment will test for deviations from $1/r^2$ in the force law, search for
violations of the EP and measure Newton's constant $G$ to one part in $10^{7}$.
This will be achieved by accurately determining the orbit of two tests masses as they interact gravitationally with
each other and with the Earth. For our purposes we shall focus on the expected value of $G$, including the contribution from the fifth force mediated by $\phi$. 

We first show that it is possible for the satellite {\it not} to have a thin shell. This is desirable in this case in order to maximize the signal. That is, we derive the conditions under which $\Delta R_{SEE}/R_{SEE} > 1$ holds for the SEE capsule. According to the current design, the satellite will have a total mass of $\sim 2000$ kg and an effective radius of $R_{SEE}\sim 2$~m (although it has cylindrical symmetry, we approximate the capsule as a sphere of equal volume for simplicity). Thus its Newtonian potential is $\Phi_{SEE} \approx 10^{-24}\approx 10^{-15}\Phi_\oplus$. Moreover, at an altitude of 1000 km, the chameleon field assumes the value $\phi(r_{SEE})\sim  \phi_G$, as seen from Eq.~(\ref{earthext}). Therefore, the condition $\Delta R_{SEE}/R_{SEE} > 1$ requires
\begin{equation}
\frac{\Delta R_\oplus}{R_\oplus} > 10^{-15}\,.
\label{noscreen}
\end{equation}
Combining this with the condition that the atmosphere have a thin shell, Eq.~(\ref{condition2}), we find that the allowed range for which the SEE satellite will {\it not} have a thin shell is
\begin{equation}
10^{-15} < \frac{\Delta R_\oplus}{R_\oplus} < 10^{-7}\,.
\label{atmGG}
\end{equation}
This constitutes a wide region in parameter space. Following the analysis of Sec.~\ref{condmass}, for a potential $V(\phi)=M^{4+n}\phi^{-n}$ with $n=1/3$ and $\beta=10$, this requires the range of interactions in the atmosphere, $m_{atm}^{-1}$, to be anywhere between $0.1\;\mu$m and 1 mm.

Therefore, if the inequality in Eq.~(\ref{noscreen}) is satisfied, the scalar field $\phi$ is only slightly perturbed by the satellite, and thus the field value within the capsule is $\phi\sim \phi_G$. Moreover, since $m_G^{-1}$ is much larger than the size of the satellite, $\phi$ behaves as a massless field. Hence, if the test bodies have mass $M_i$ and coupling $\beta_i$, $i=1,2$, they will experience a total force, 
gravitational plus $\phi$-mediated, given by
\begin{equation}
|\vec{F}| = \frac{GM_1M_2}{r^2}\left(1+2\beta_1\beta_2\right)\,,
\label{F12}
\end{equation}
where we have restored Newton's constant $G=(8\pi M_{Pl}^2)^{-1}$.
This implies, therefore, that the SEE satellite will measure an effective Newton's constant,
$G_{eff} \equiv G (1 +2\beta_1\beta_2)$, {\it which differs by order unity from Newton's constant measured on Earth}.

Since this is an important prediction, let us summarize how it was obtained. First of all, the key ingredient is that the
SEE experiment is performed in orbit, where the range of $\phi$-mediated interactions is greater than 10 AU. Thus, small
perturbations in $\phi$ are essentially massless on the scale of the capsule, and therefore the force law between
two test bodies is proportional to $1/r^2$ to a very good approximation. A second crucial condition is that there is no
thin-shell effect within the satellite, as expressed in Eq.~(\ref{noscreen}). This ensures that there is no suppression
of the fifth force, and the coupling strength is therefore of order $\beta^2$. Hence, for 
$\beta$ of order unity, the effective Newton's constant receives large corrections from the fifth force mediated
by $\phi$. As mentioned earlier, such an outcome would constitute strong evidence that a mechanism of the kind
we are proposing is at play in Nature.

Our model also has surprising implications for satellite experiments that aim at measuring
EP violations in orbit, such as MICROSCOPE, GG and STEP. These experiments will attempt to measure a difference in free-fall acceleration between two concentric cylinders of different composition with expected sensitivity of one part in $10^{15}$, $10^{17}$ and $10^{18}$, respectively. 

We focus on the STEP satellite for concreteness. Although Eq.~(\ref{noscreen}) pertains to the SEE capsule, a similar condition is obtained 
for STEP, since its physical characteristics are not too different from those of SEE. Hence, there is no thin-shell effect if Eq.~(\ref{noscreen}) holds, and the $\phi$ profile within the satellite is well-approximated by Eq.~(\ref{earthext}). It is then straightforward to show that the E$\ddot{{\rm o}}$tv$\ddot{{\rm o}}$s parameter for Be and Nb test cylinders, as appropriate for STEP, is given by
\begin{equation}
\eta \approx 10^{-4}\beta^2 \frac{\Delta R_\oplus}{R_\oplus}\,. 
\label{etaGG}
\end{equation}
Combining with Eq.~(\ref{atmGG}), we find the allowed range
\begin{equation}
\beta^2\cdot 10^{-19}< \eta < \beta^2\cdot 10^{-11}\,,
\label{range}
\end{equation}
which falls almost entirely within the range of sensitivity of STEP. Moreover, we see that $\beta$ can be larger than $10^{-13}$, the current bound from experiments performed
on Earth~\cite{eotwash}. In other words, it is possible that MICROSCOPE, GG and STEP will measure violations of EP stronger than currently thought to be allowed by laboratory measurements.
This is another striking manifestation of the fact that scalar field dynamics in our model are very different on Earth than in orbit.

\section{Discussion} \label{conclu}

In this paper, we have presented a novel scenario in which both the strength and the range of interactions mediated by a scalar field $\phi$ depend sensitively on the surrounding environment. In a region of high density, such as on Earth, the mass of the scalar field is sufficiently large, typically of order 1 ${\rm mm}^{-1}$, to evade constraints on EP violation from laboratory experiments. Meanwhile, the field can be essentially free on solar system scales, with a typical Compton wavelength of 100 AU. 

We have argued that our scenario satisfies all existing bounds from ground-based and solar-system tests of gravity. It will be left for future work to study our model in the context of strongly gravitating systems, such as black holes~\cite{trodden}. Implications for cosmology will be analyzed in a separate paper~\cite{cosmo}, where we will show in detail that chameleons are consistent with cosmological constraints on the existence of non-minimally coupled scalars, such as the bound on the time-variation of $G$ from nucleosynthesis, for example. Moreover, we will describe how the field dynamically reaches the minimum of the effective potential, an element that was assumed {\it a priori} in the present work, and show how our scenario naturally provides a solution to the old moduli problem~\cite{cough}.

The striking difference in interaction range in the solar system versus on Earth can lead to unexpected outcomes for experiments that test gravity in space. It was argued that the SEE Project could measure an effective Newton's constant drastically different than that on Earth. Meanwhile, the MICROSCOPE, GG and STEP satellites could detect violations of the EP stronger than currently allowed by ground-based experiments. Either outcome would effectively constitute a proof of the existence of chameleons. More importantly, these possible surprises for space experiments will hopefully strengthen the case for launching these satellites.

Also, the door is open for conceiving of new table-top experiments to probe for the existence of chameleons, for instance by exploiting the fact that the properties of chameleons are sensitive to the surrounding matter density. For example, one could imagine doing atomic physics experiments inside or in the vicinity of a massive oscillating shell of matter. These oscillations would induce a time-dependence in the profile of the chameleon and thus in the frequency of emission lines~\cite{gibbonscomment}. It would also be interesting to investigate the behavior of the field at intermediate altitude, for instance at about 40 km where balloon experiments take place. This would require a more realistic modeling of the atmosphere than presented here. Such an analysis could reveal that it may be possible to detect chameleons by performing gravity experiments in balloons~\cite{steincomment}.

We thank N. Arkani-Hamed, J.R. Bond, R. Brandenberger, P. Brax, C. van de Bruck, S. Carroll, T. Damour, A.-C. Davis, G. Esposito-Far\`ese, G. Gibbons, B. Greene, D. Kabat, A. Lukas, J. Murugan, B.A. Ovrut, M. Parikh, S.-J. Rey, K. Schaalm, C.L. Steinhardt, N. Turok, C.M. Will, T. Wiseman, and especially N. Kaloper and P.J. Steinhardt for insightful discussions.
This work was supported by the Columbia University Academic Quality Fund, the Ohrstrom Foundation (JK), DOE grant DE-FG02-92ER40699 and the University of Cape Town (AW).

\end{document}